\documentclass[
 reprint,
 amsmath,amssymb,amsfonts,
 aps,
prstab,
]{revtex4-2}

\usepackage{graphicx, subcaption}
\usepackage{dcolumn}
\usepackage{bm}
\usepackage[allcolors=blue]{hyperref}

\usepackage{siunitx, pgf, import}
\usepackage{float}
\usepackage[super]{nth}

\usepackage{float}
\makeatletter
\let\newfloat\newfloat@ltx
\makeatother

\usepackage{algorithm}
\usepackage[noend]{algpseudocode}

\usepackage{tikz}
\usetikzlibrary{arrows,positioning}
\tikzset{
    >=stealth',
    punkt/.style={
           rectangle,
           rounded corners,
           draw=black, very thick,
           text width=3.5em,
           minimum height=2em,
           text centered},
    pil/.style={
           ->,
           thick,
           shorten <=2pt,
           shorten >=2pt}
}
\usepackage{wasysym}

\begin{document}

\preprint{APS/123-QED}

\title{Towards automatic setup of \SI{18}{MeV} electron beamline
using machine learning}

\author{Francesco Maria Velotti}
\email{francesco.maria.velotti@cern.ch}
\author{Brennan Goddard}
\author{Verena Kain}
\author{Rebecca Ramjiawan}
\author{Giovanni Zevi Della Porta}
\affiliation{CERN, Geneva, CH}
\author{Simon Hirlaender}
\affiliation{University of Salzburg, Kapitelgasse 4/6, 5020 Salzburg, Austria\\\\}

\date{\today}


\begin{abstract}

 To improve the performance-critical stability and brightness of the electron
 bunch at injection into the proton-driven plasma wakefield at AWAKE, automation approaches
 based on unsupervised Machine Learning (ML) were developed and deployed.
 Numerical optimisers were tested together with different model-free
 reinforcement learning (RL) agents. To aid hyper-parameter selection, a full
 synthetic model of the beamline was constructed using a variational
 auto-encoder trained to generate surrogate data from equipment settings. This
 paper introduces the AWAKE electron beamline and describes the results obtained
 with the different ML approaches, including automatic unsupervised feature
 extraction from images using computer vision. The prospects for operational
 deployment and wider applicability are discussed.

\end{abstract}
\maketitle

\section{Introduction and Motivation}

\begin{figure*}[ht] \centering\includegraphics[width=0.8\linewidth, trim={0mm
    0mm 0mm 0mm}]{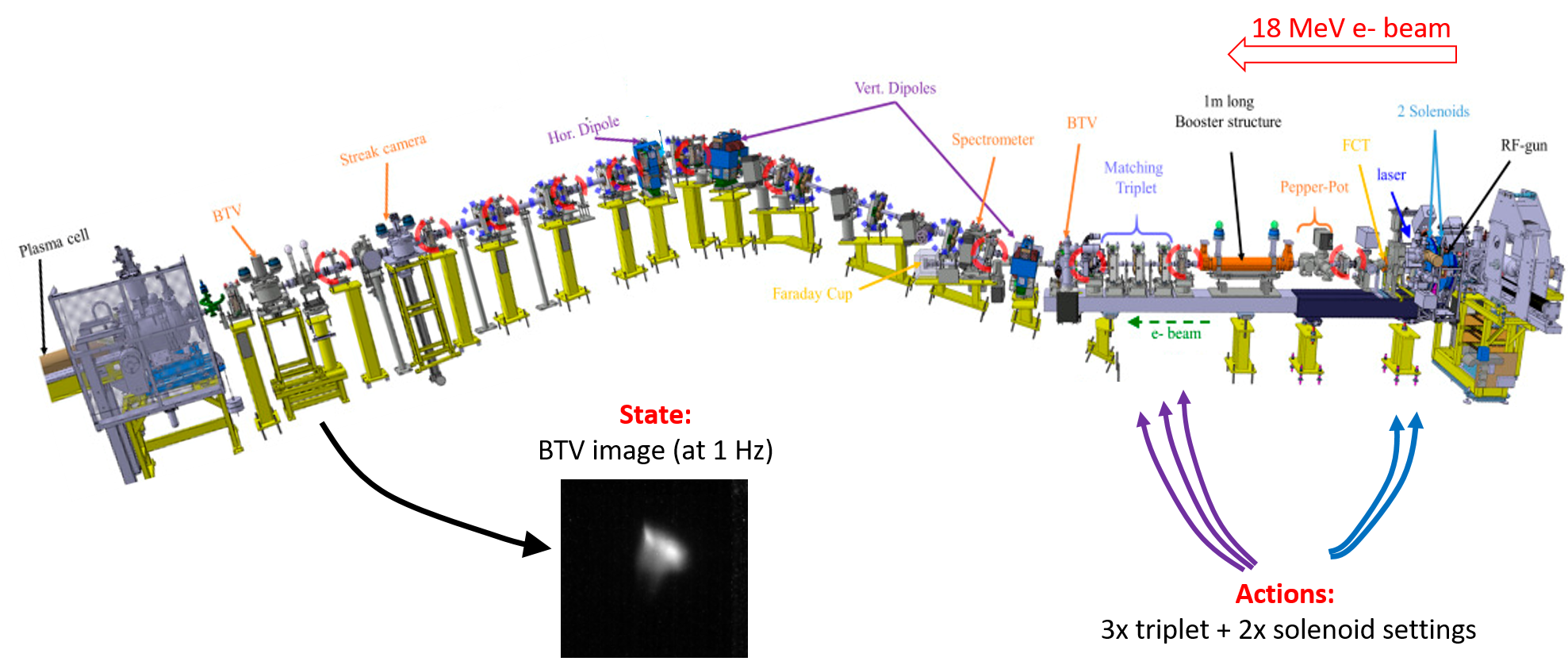} \caption{AWAKE beamline showing
location of the matching devices (actions) and the observation BTV.}
\label{fig:AWaKE_beamline} \end{figure*}

The AWAKE experiment~\cite{AWAKE01} at CERN's Super Proton Synchrotron (SPS) uses proton-driven plasma
wake-fields to accelerate an 18~MeV electron witness bunch to about 2~GeV over a
distance of 10~m. Efficient capture and acceleration relies on precise delivery
of a dense low-energy e$^-$ bunch to the correct location in space and time in
the plasma. The e$^-$ beam brightness (intensity divided by transverse beam size) and
position are therefore critical to the performance of the overall facility, as
evidenced by the experiments performed in 2018 \cite{AWAKE2018}. The work reported in this
paper complements the encouraging initial results obtained for multi-objective
optimisation of trajectory and emittance~\cite{alex_multi}.

The low energy \SI{18}{MeV/c} e$^-$ beamline requires time-consuming and
frequent optimisation, given its high sensitivity to initial conditions, to
equipment settings and to the bunch momentum distribution, as well as inherently
less predictable environmental effects like temperature, magnetic history and
the pulsing of the adjacent \SI{400}{GeV/c} proton beamline. The
commissioning of the e$^-$ beamline highlighted the criticality of the magnetic
element modelling and of the incoming beam energy jitter~\cite{eline_comm}.

Some of the contributions to beam quality degradation are completely random or
uncontrollable, with timescales ranging from seconds to hours or even days.
Time-consuming and sometimes non-reproducible manual tuning of the e$^-$ source
and beamline parameters were needed to satisfy the experiment requirements.

The main fluctuations are observed on the transverse beam quality and are
believed to be caused by chromatic aberrations and optical mismatch at the
injection point. Problems of this type for low-energy lines are known
\cite{psi, swissfel}, although  large variations usually also affect the longitudinal
beam delivery. A similar multi-objective optimisation is needed, e.g. position,
angle, emittance and charge delivered, with a large number of free tuning
parameters.

To reduce the time and personpower effort needed for setting up, and to improve the
stability and also potentially the absolute performance reach, model-free
ML automation approaches were investigated.
Acting on the initial matching
triplet and the low-energy solenoids, the beam brightness was optimised, using as
observation the image of the beam on a beam monitoring screen (BTV~\cite{btv}) at the entrance of the plasma cell,
Fig.~\ref{fig:AWaKE_beamline}. To test and deploy the different types of
optimisation agents, an interface to the SPS control system was used together
with the generic OpenAI Gym~\cite{openai} environment framework for the ML tasks. A surrogate
model using computer vision in a Variational Autoencoder (VAE) trained on the machine data was an
important part of the work, allowing fast \textit{in-silico}  testing and tuning
of different algorithms and approaches without beam time.

For the optimisation of the beam brightness, the two different approaches investigated
were numerical optimisers and reinforcement learning. For the numerical
optimisers, by varying the equipment parameters the algorithm aims to maximise
or minimise an objective function calculated from the BTV image, which it must
perform each time it is used. For RL, the agent aims to learn the response of
the system during a training phase, such that it can quickly move to the optimum
in the subsequent deployment.

The relevant performance metrics for both types of approach are the sample
efficiency (number of interactions needed with the machine for the algorithm to
converge) and the final beam brightness achieved. The RL agents, based on the Markov assumption, have the
advantage of not needing to repeat the exploration phase, once the underlying
dynamics have been learned, but unlike the optimisers will not perform well on
subsequent deployment if the underlying dynamics of the system are non stationary.

A key requirement for RL is that an appropriate state information can be provided
to the agent. In our experiments, we tested both \textit{explicit} state
encoding, using the output of analytical Gaussian image fits, and \textit{
implicit} state (or feature extraction) encoding, where the encoder from a trained VAE gives a
representation of the image in a low-dimensional $\mathbb{Z}$ latent space,
which was then used directly as implicit state information for the RL agent. This automatic
unsupervised feature extraction could be critical for RL applications where
 explicit state feature description and extraction is difficult (very high dimensional problems) or impossible,
for instance in the observation of Schottky spectra.

To facilitate hyper-parameter optimisation, agent selection and investigate
transfer learning, the decoder of the trained VAE was also used to generate a
full synthetic model of the system. This synthetic model is able to encode and
decode images to and from a latent space $\mathbb{Z}$ using an additional
predictor neural network to ensure the correspondence between $\mathbb{Z}$ with
the equipment setting configuration $\mathbb{C}$. In this way, it can replace
the real beamline to help tune and test any algorithms.

This paper introduces the AWAKE e$^-$ beamline with its operational challenges,
and explains the technique for matching for maximum beam brightness at the
injection point. The methodology for the implementation of the different
optimisers and RL agents is presented, together with the VAE. The construction
of the synthetic model is briefly described. The performances of the different
approaches deployed on AWAKE are compared, including comparison with the
synthetic model results. Technical aspects such as implicit versus explicit
state representation are addressed, including a technique the authors developed for
overcoming the inherent difficulty with RL reward shaping which simplified and
stabilised training and improved overall sample efficiency. Finally, future work
and the prospects for operational deployment and wider applicability are
discussed.

\subsection{AWAKE electron transfer line and optics}

The AWAKE experiment uses a 400~GeV/c proton transfer line to transport the
drive beam from the SPS to the plasma cell. The 18~MeV e$^-$ beam is produced in
a side-gallery and needs to be fed into the same plasma cell with high delivery
precision. An initial S-band RF photo injector produces a \SI{200}{pC} e$^{-}$
beam at around \SI{5}{MeV/c} which is then accelerated in  a travelling wave
accelerating structure up to about \SI{18}{MeV/c}~\cite{e_source}. Two low
energy solenoids are used just after the photo injector to focus the beam
inside the accelerating structure and hand it over to the transfer line. The
latter is equipped with an initial and final matching quadrupole triplets to
ensure losses transport and final focus at the plasma cell entrance.

\begin{figure}[h]
    \centering\includegraphics[width=0.45\textwidth]{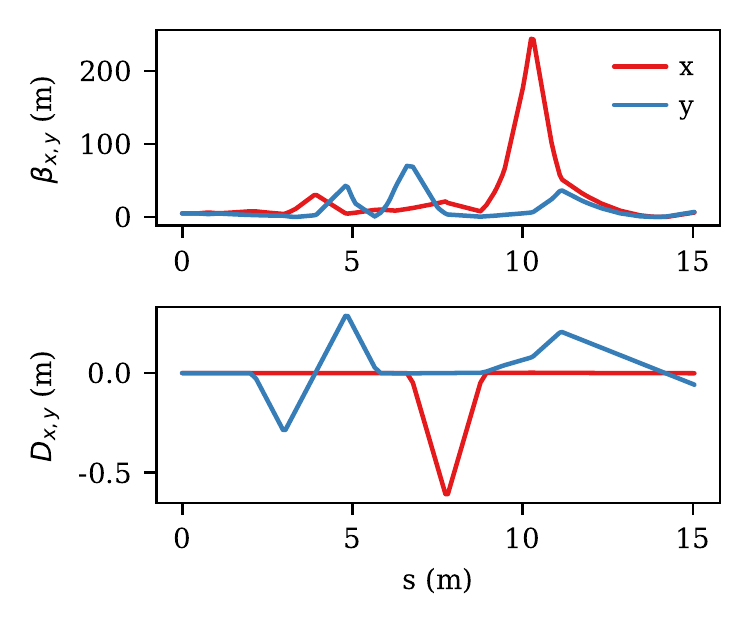}
    \caption{Electron beamline linear optics functions. Top: beta functions for both
    horizontal and vertical plane. Bottom: dispersion functions.}
    \label{fig:optics} \end{figure}

The existing tunnel geometry was an important constraint on the optics design of
the e$^{-}$ line. In  Fig.~\ref{fig:optics} the optics of the e$^{-}$ beamline
is shown. It comprises two achromatic sections: one vertical dog-leg and one
\SI{60}{\degree} horizontal bend to go from the RF gun to the plasma cell, as
well as matching elements. Due to the difference in the vertical slope between
the tunnel of the RF gun and the plasma cell, the vertical dispersion is matched
to zero locally at the merging point but with a finite dispersion angle.


In order to ensure capture of the injected e$^-$ in the plasma accelerating
structure, the beam size at the entrance of the plasma cell has to be
\SI{250}{\micro m} in both planes, with as high intensity as possible. Due to
the strong bends and quadrupoles in the beamline, the chromatic aberrations
significantly degrade the deliverable beam quality~\cite{eline_comm}. This is
accentuated by the shot-to-shot momentum jitter observed from the e$^{-}$
source.

The AWAKE e$^-$ line commissioning \cite{eline_comm} showed that most of the
non-reproducibility is due to variations in the beam from the source, which
changes the matching needed to the target. This translates into lengthy setting
up to produce the required beam parameters, which must be repeated every time
the source is restarted.

\section{Optics matching}

The beam brightness and position need to be controlled at the injection point.
The dependence on high order aberration of the optics plus the variations in the
initial conditions excludes purely analytical matching of the target beam size
$\sigma^*$ using only linear optics, as a global solution. In fact, analytical
matching is possible only if the initial conditions are close enough to the
design.

The last beam screen (BTV) is located \SI{0.8}{m} upstream of the injection point,
as installation inside the plasma cell was not possible. The spot brightness can
only be maximised at this screen. An optics trim then needs to be applied to
move the optics waist to the required location. Accurate knowledge of the beam
optics functions is thus fundamental.

With low energy e$^-$ the usage of multiple screens for single-shot optics
measurement is impossible as the beam is completely disrupted by the screen. For
the AWAKE e$^{-}$ line an ad-hoc measurement optics was developed which presents
a global minimum at a specific longitudinal location. The brightness is then
maximised on the final screen, using only the upstream triplet. The global
minimum is moved from this screen to the injection point using only 4
quadrupoles in the final part of the line, and leaving the initial matching
quadrupole strengths unchanged.

In theory the dispersive contribution should be taken into account, but this is
not so important for our specific application, as the initial dispersion is very
close to zero. If the beam is well centred in the initial
triplet~\cite{eline_comm}, no significant contribution from the variation of the
first three quadrupoles is expected to the dispersion functions.

The displacement of the focal point was fully tested in simulations. Although not yet
deployed experimentally, the results and methods shown in this paper are not linked
to the success of this methodology. For the future, the installation of a BTV in
the plasma volume at the injection point is under investigation, which would
remove the need for this step.

\begin{figure*}[ht]
\centering\includegraphics[width=\linewidth]{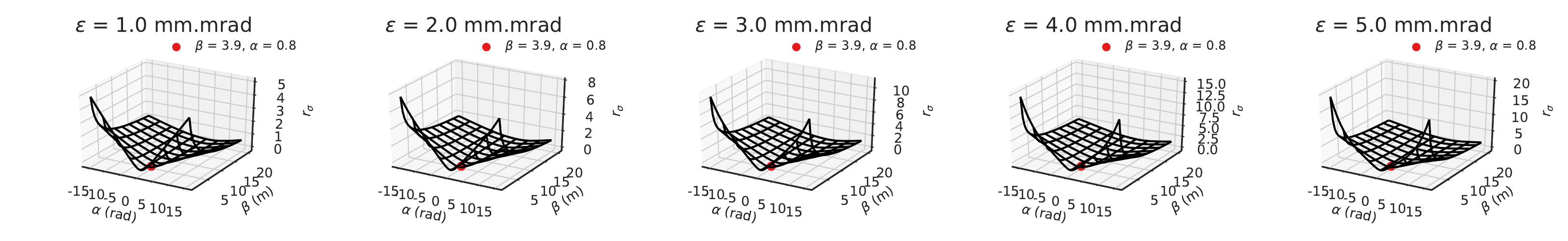}
\caption{Scan of sensitivity of matched beam spot-size to initial optics
conditions, for different emittances.} \label{fig:r_vs_ini} \end{figure*}

The beam quality at the end of the line is also optimised with the low energy
solenoids before the accelerating structure, which help to minimise the
emittance produced. This can be achieved using the following penalty function
$r_{\sigma}$,
\begin{equation} r_{\sigma} = \sqrt{(\sigma_x - \sigma_x^*)^2 + (\sigma_y -
\sigma_y^*)^2} \end{equation}.

choosing $\sigma^*$ using the lowest achievable emittance from
the gun (as previously measured, i.e. \SI{0.9}{mm.mrad}). This ensures that the
target function has a single global minimum in the 5-dimensional action space.
The validity of this approach was tested in simulation using a model including
known non-linear effects. Figure~\ref{fig:r_vs_ini} shows that the evolution of $r_{\sigma}$ as a function
of beam initial conditions, i.e. $(\beta, \alpha)_{x, y}$, and for
different initial emittances. As the figure of merit is in each case a convex function with a
clear minimum (red dot), $r_{\sigma}$ can be incorporated into a target function to
ensure the matching of the beam produced from the source to the transfer line.
The same minimum location is also found in initial optics space for different
emittances, which shows that the approach should be insensitive to emittance
variation.

\section{Machine Learning framework and methodology}

Thanks to the high repetition rate of the AWAKE e$^-$ source, a large set of
machine learning algorithms can be explored with the matching approach described
above. The python Open AI gym environment framework was used as a
standard, with the pyJAPC~\cite{pyJAPC} library for interfacing between the ML
algorithms and the CERN control system. The TensorFlow (version 1.14) back-end
was used for the machine vision and VAE. Different numerical optimisers and
Stable Baselines~\cite{stablebaselines} RL agents were explored for both
explicit and implicit state representation. BOBYQA~\cite{bobyqa} and Twin
Delayed Deep Deterministic policy gradient algorithm (TD3)~\cite{td3} gave the
best results in the two classes.

\subsection{Action space}

For all synthetic and machine tests described, the Action space consisted of the
2 solenoid and the 3 quadrupole currents, as indicated in
Fig.~\ref{fig:AWaKE_beamline}. In the OpenAI gym agents, the values are
normalised to a range of $\pm1$, while in reality the solenoid currents are
0-400~A and the quadrupoles -100 to 100~A. The Action space is
therefore bounded to the physical limits of the different circuits, such that no
other limits or penalties are needed for the agents. Although the line runs at
10~Hz, some time is needed between setting and acquiring which is dealt with in
the generic OpenAI environment, which limits the rate at which scans can be made
to about 0.5~Hz.

\subsection{Observation}

The observation is the BTV image,  Fig.~\ref{fig:AWaKE_beamline}, from which we
derive both the single objective function (brightness) and the state for the RL
agents. The BTV image provides a 256$\times$256 pixel 8-bit grey-scale array, from
which beam profiles, intensity and more complex information can be extracted at
1~Hz. The images were down-sampled to 128x128 pixels, which was
the dimensionality for the VAE encoder and decoder used in the synthetic model.

\subsection{Optimiser objective function}

All results reported here were obtained optimising a single objective function.
Studies made using multi-objective optimisation with an extremum-seeking
optimiser have been reported separately \cite{alex_multi}.

The optimiser tries to minimise the objective function, which should therefore
be large and positive when far from the optimum, and small when close to the
ideal solution. The objective function used for the optimisers was defined as
composition of two contributions:

\begin{equation}
r_i = \frac{1}{i_0}\sum_{j,k} a_{jk} - i_0
\label{eq:i0}
\end{equation}

\begin{equation}
r_{\sigma} = r_{0} - \frac{1}{r_{max}} \sqrt{(\sigma_x -
\sigma_x^*)^2 + (\sigma_y - \sigma_y^*)^2}
\label{eq:rs}
\end{equation}

where $\sum_{i,j} a_{ij}$ is the measured sum of all pixel values and
$\sigma_x^*$ and $\sigma_y^*$ were both set to 0.1~mm to represent a target
minimum beam size, $i_0$ was set at a numerical value of $1.3\times10^6$ slightly above the
maximum ever recorded sum of pixels; $r_0$ and $r_{max}$ were set to 0.25~mm and
3.0~mm respectively, as minimum and maximum spot sizes. The two contributions are then
put together to for the actual penalty function for the optimiser as:

\begin{equation}
    r_o = -1 [r_{\sigma} \alpha_s + r_i (1 - \alpha_s)]
\label{eq:sum-r}
\end{equation}

where $\alpha_s$ represent the weight for the beam size contribution. It was empirically found
that 20\% weighted contribution from the image
amplitude $r_a$ over all pixels and 80\% from the beam size $r_{\sigma}$ is a good compromise to
achieve the desired beam parameters. The function is designed
to encourage simultaneous high intensity and small beam size.

\subsection{Generative VAE for synthetic model}

A Variational Autoencoder (VAE~\cite{VAE}) based on computer vision convolutional neural
networks was used to generate synthetic BTV images from the real AWAKE data in
an unsupervised manner. This was then used both for state encoding from the BTV
images, as well as building a synthetic model (digital twin) of the AWAKE
beamline. The basic autoencoder (AE) is a pair of neural networks consisting of
an encoder, an information bottleneck, and a decoder. The loss function is built
of two parts: the reconstruction accuracy which measures how close the decoded
data is to the original data, and a divergence term which measures how the
information contained in the latent encoding differs from a Gaussian distribution. The pair of
networks try to reconstruct the original data as accurately as possible, passing
through the low-dimensional information bottleneck.

\begin{figure}[ht]
\centering\includegraphics[width=0.5\textwidth]{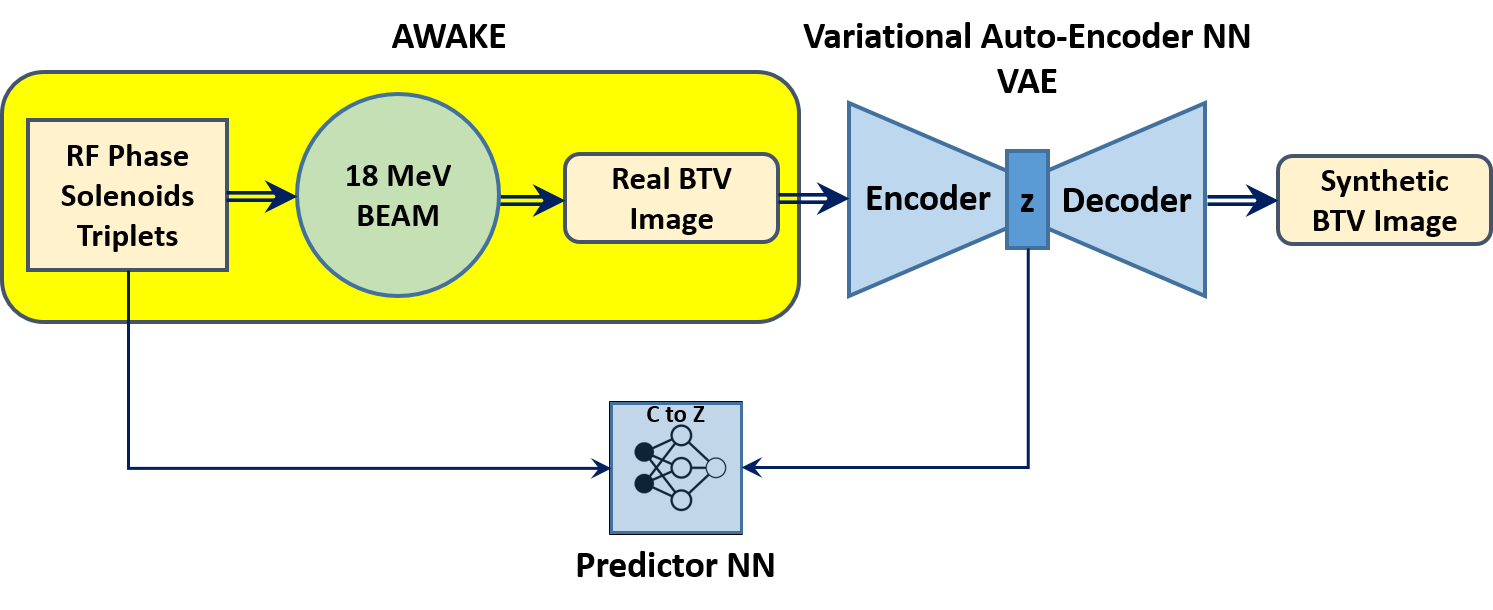}
\caption{Architecture for training VAE and Predictor networks used for AWAKE
synthetic model and RL encoding. } \label{fig:vae}
\end{figure}

\begin{figure}[ht]
    \centering\includegraphics[width=0.35\textwidth]{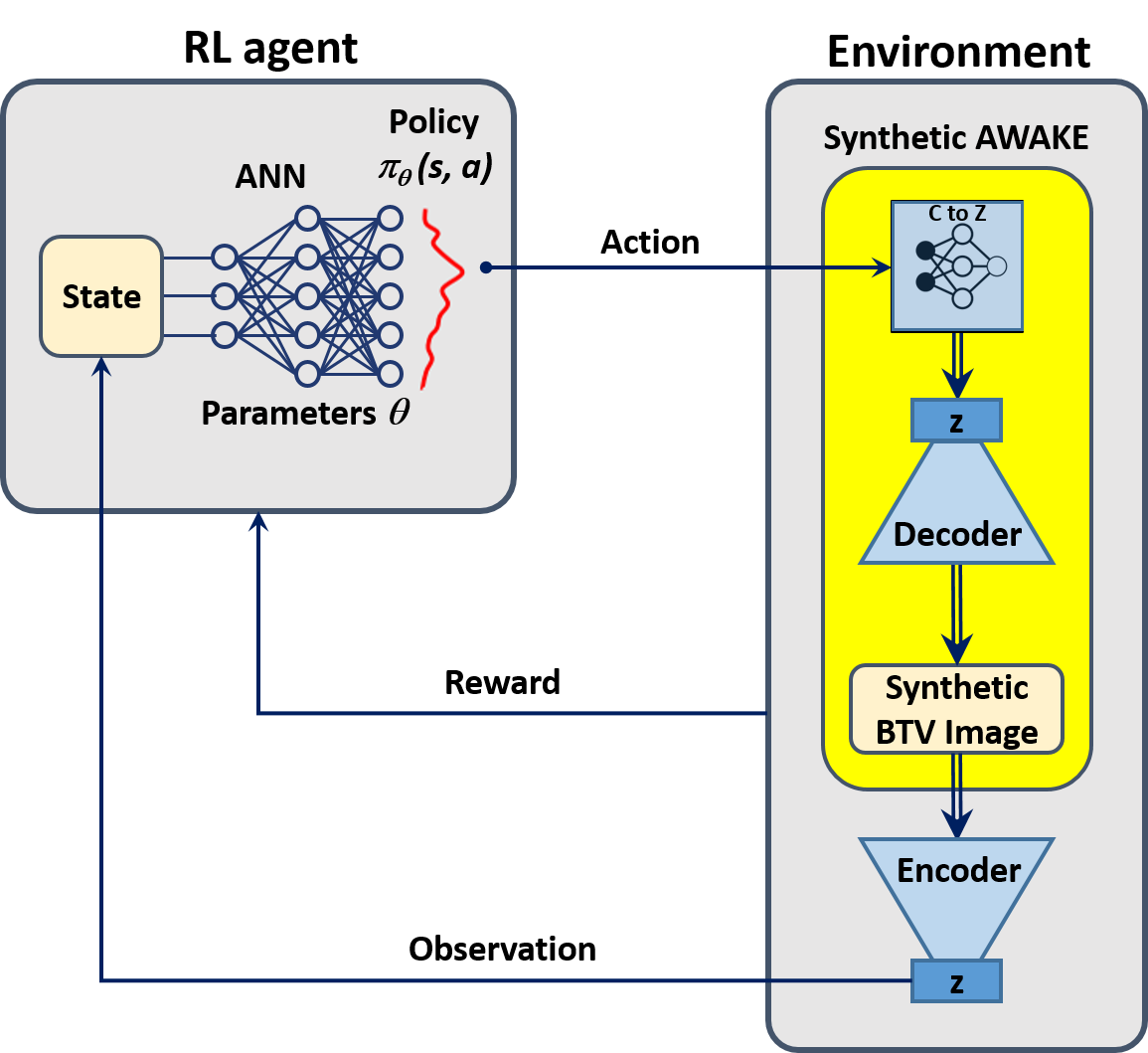}
    \caption{Use of AWAKE synthetic model for RL testing. Note that the decoder
    and encoder networks do not need to come from the same VAE model.}
    \label{fig:synthetic} \end{figure}

The VAE uses an additional random term added to the
encoded latent space coordinate. Even with limited discrete training data this
has the effect of producing a continuous variation of encodings in the latent
space, ideal for state variables which we expect to be continuous with changes
in the actions for our system. In our experiments we tried different VAE
flavours, settling for the $\beta$-VAE with loss function of the form:

\begin{equation}
\begin{split}
    L(\theta, \phi, \beta) = - E_{z\approx q_\phi (\mathbf{z}  \mid \mathbf{x})}
    \log p_\theta (\mathbf{x}  \mid \mathbf{z}) + \\
    \beta D_{KL}(q_\phi (\mathbf{z}|\mathbf{x}) \mid \mid p_\theta(\mathbf{z}))
\end{split}
\end{equation}

as from~\cite{betaVAE}, where the first term represents the reconstruction loss and the
second one is the Kullback-Leibler divergence (KL) which pushes the probability distribution
of decoder and encoder to be as similar as possible to a Gaussian distribution. The KL term
is weighted with the $\beta$ parameter which can be considered an additional hyper-parameter to
choose (examples of produced images are shown in Annex Fig.~\ref{fig:vae_images_z}).
To complete the full synthetic model, a densely connected neural network was used as
surrogate model to make the correspondence between the (labelled) Action space
and the latent space $\mathbb{Z}$ encoding. The overall architecture for
training the VAE and Predictor is shown in Fig.~\ref{fig:vae}, with the
synthetic model shown in Fig.~\ref{fig:synthetic}.

These networks were trained on machine data, from a grid-scan made in the 5-D
action space. This was a lengthy one-off process needing the accumulation of
some \num{1500} valid images to train the VAE, but allowed the efficient and
comprehensive off-line training and hyper-parameter optimisation for comparison
of different agents, objective functions, state encoding and reward shaping for
the different agents investigated.

The lengthy grid search was made to train the surrogate model but could theoretically
be used to find the optimal working point. But it would not be a viable
method as the beamline input conditions vary quite significantly from day to day
(which also impacted the suitability of the single RL agent approach, as addressed
in the discussion).

\subsection{RL state space}

RL agents need a state space based on the observation, as well as an action
space which changes the state. Two approaches for state representation were
investigated. For explicit state extraction, the parameters $\sigma_x$,
$\sigma_y$, $\mu_x$ and $\mu_y$ were obtained from numerical fitting of a 2D
Gaussian to the BTV image and the intensity extracted from the pixel sum. For
implicit state extraction from the image we used the computer vision encoder
network from a trained VAE and fed the latent vector $\mathbb{Z}$ directly to the RL agent
as the state representation. In this case it is important to note that
the elements of the state vector do not
correspond directly to individual physics parameters.
This automatic unsupervised feature extraction could be essential for
problems where the explicit extraction of state features is difficult or
impossible, or where some hidden state features are to be expected. Given that
some of the profiles obtained from the AWAKE BTV are highly non-Gaussian (see
Annex Fig.~\ref{fig:zreconstruction}), it was hoped that this implicit approach would
help more completely capture the underlying image dynamics.

Since most of the information encoded in a low-dimensional latent space
corresponds to the position of the beam spot, we experimented testing a centring algorithm to
produce the correct sized data array centred about the
brightest part of the image. The results were rather similar and we finally ran
almost all experiments without explicitly correcting for the beam position on the
screen. This was valid in the context of optimising for the beam
brightness - separate studies have shown that the beam position stabilisation
can be treated separately~\cite{alex_multi}, using dipole correctors and
position monitors.

\begin{figure}[ht]
    \centering\includegraphics[width=0.35\textwidth]{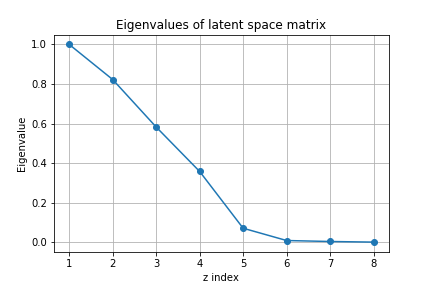}
    \caption{Eigen modes of latent space matrix for encoded images with
    $\mathbb{Z}$-dimension of 8. It can be seen that only 5 significantly
independent image modes exist,} \label{fig:eigenModes} \end{figure}

An investigation of VAE architecture and hyper-parameters was made to achieve
stable results. Different flavours of VAE were tried in attempts to produce a
more disentangled latent space description, eventually we used a $\beta$-VAE
flavour \cite{betaVAE}. The dimensionality of the $\mathbb{Z}$ space was also
studied in terms of suitability for RL state encoding, since a larger value
allows more accurate reconstruction, but would intuitively seem likely to complicate the
encoding challenge for the RL agent. One useful metric was the eigenmode
decomposition of the latent space matrix for all encoded images, \ref{fig:eigenModes},
which allowed
us to see how many of the latent dimensions were actually encoding independent
information, see for example Fig.~\ref{fig:eigenModes}.

Tests on the synthetic AWAKE model confirmed that RL agents using a state
encoding dimension of 64 or 512 failed to converge. We therefore fixed the
latent space dimensionality at 5.

\subsection{RL reward}

The RL agent needs a reward which is provided after each action. One important
aspect is that the agent aims to maximise the reward over the learning process.
Since we are interested in finding a sample-efficient solution which takes as
few action steps as possible, our reward function needs to be large and negative
when far from the optimum solution, and small but still negative when close to
the optimum.

We found that an appropriate choice of reward function per iteration $r_i$ was a
very important factor in stable RL agent performance. Our function was
constructed to be always negative for realistic observation parameters, and to
have a maximum of around 0. This was done combining Eq.~\ref{eq:rs} and Eq.~\ref{eq:i0},
as done in Eq.~\ref{eq:sum-r} but multiplying it by -1 to obtain suitable reward for
our agent:
\begin{equation}
    r_i \equiv -r_o
\end{equation}

Importantly, the two
separate contributions were clipped in the range $[-1,0]$.

Another aspect that was investigated was the ending-episode reward: this is usually
a large positive number which would significantly increase the cumulative reward
along the whole episode. It was experimented with and without and found that the
difference in terms of number of machine iterations needed for full training was
negligible. To compare different techniques to represent the observation space
though, the end-episode reward was found to help the metric chosen to classify
them, and hence only in this particular case, a positive reward of 20 (chosen
arbitrarily large) was assigned to the agent at successful completion of an
episode.

\subsection{RL episode termination and reward dangling}

The correct termination of the RL episode was also a critical factor in the
stable performance. For this, we introduced a \textit{reward target} $r_t$,
which when achieved terminated the episode. This introduced a new problem, since
the correct setting of this threshold value then also turned out to be an
important hyper-parameter. Too low (easy) and the agent would not achieve a good
performance, while too high (hard) and the agent would fail to train. Since the
reward at the start of the episodes varied unpredictably, and also since the
final achievable performance varied depending on the specific run conditions, we
needed to develop an automatic way of setting $r_t$. We opted for \textit{reward
dangling}, where we split the RL agent training into two parts - in the first
part, $r_t$ was fixed at a very low easy value, typically -0.5. A pseudocode
of the procedure is detailed below:

\begin{algorithm}
\caption{Reward dangling}
\begin{algorithmic}
\Procedure{reward dangling}{$\alpha, \gamma$}
\State $\alpha = 0.1$
\State $\gamma = 0.99$
\State $r_t \gets -0.5$
\While{$i_e < N_{e, max}$}
\State $r_f \gets$ Run-episode-training$(i_e, r_t)$
    \If{$r_f > r_t$}
        \State $r_t \gets r_t * \gamma$
    \EndIf
    \State $i_e \gets i_e + 1$

\EndWhile
\State $r_t \gets r_t * (1 + \alpha)$
\State trained-agent $\gets$ Run-training$(r_t)$
\EndProcedure
\end{algorithmic}
\end{algorithm}

Every time that
an episode was successfully concluded, $r_t$ was then increased slightly (by
multiplying by a factor $\gamma$, typically 0.99). The training then got slightly more
difficult, until at some stage with a high $r_t$ the agent failed to train, or
takes many iterations.

The training was then repeated with a fixed value of $r_t$, set at $1 + \alpha$ times of the
final value of $r_t$ during the dangling phase. With this approach, we observed
stable results despite variations in the final value of the reward per episode.

\begin{figure*}[ht]
    \centering\includegraphics[width=0.95\linewidth]{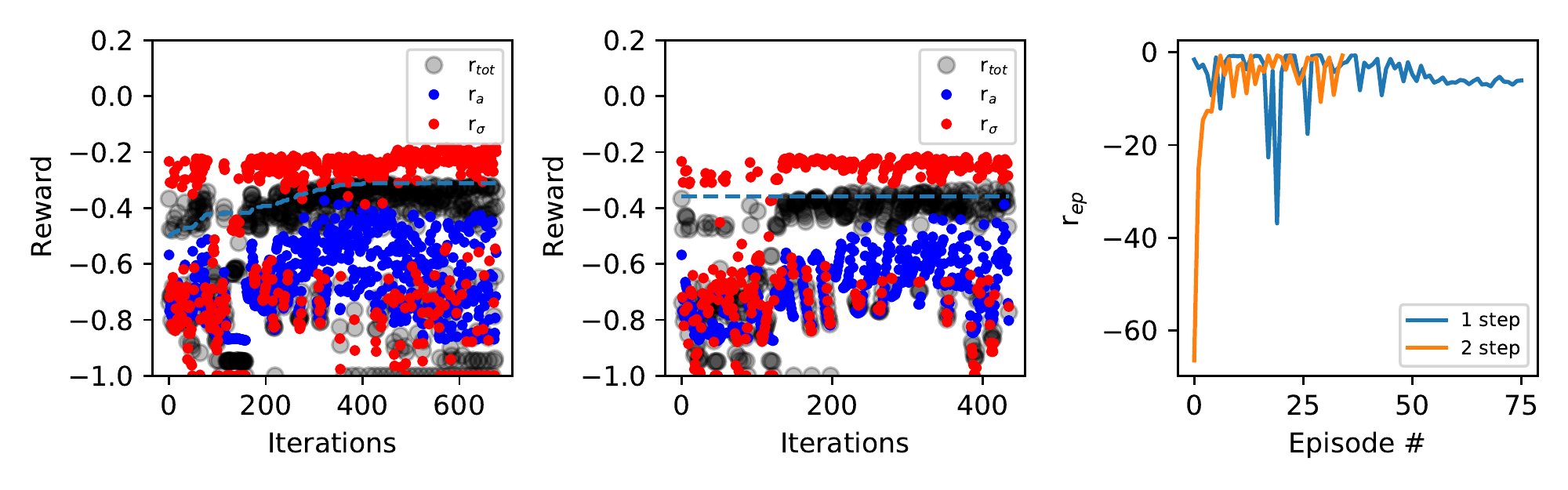}
\caption{Training of TD3 agent on synthetic AWAKE model. The first figure shows the
evolution of the total reward (black) and its two components as a function of
the iteration with the environment during the first part of the reward dangling
algorithm. The evolution of the reward threshold is shown in blue dashed line.
The middle figure shows the evolution of the total reward per episode for the
training of the agent with the chosen reward threshold. The right figures shows the evolution
of the cumulative reward as a function of the episode during the two stages of training of the
reward dangling algorithm.}
\label{fig:synthetic-results}
\end{figure*}

A final validation run with the trained agent was then used to determine the
performance, starting in a random configuration in the action space. An example
of the full reward dangling technique tested on the synthetic AWAKE is shown in
Fig.~\ref{fig:synthetic-results}.

\subsection{VAE encoders trained on synthetic image data}

One of the drawbacks of training the VAE on the real AWAKE machine is the time
needed to acquire the data. Since the state encoding is implicit, we reasoned
that training an encoder to respond to artificially-generated images with
similar spatial variations could remove the need for this step. We prepared
three different data sets.

Firstly a set of about 9'500 real BTV images with distinct action parameter
settings were filtered from all 30'000 or so 2019 measurements using Isolation
Forest regression - this was necessary as the source could trip randomly during
the different data taking campaigns and these anomalous data had to be removed.
These were re-scaled to 128 $\times$ 128 pixels, divided into
6 sub-datasets of 1'588 images and normalised to the range $[0,1]$ using the
min-max pixel values of the first sub-dataset.

A fully synthetic dataset of 1'588 images was also prepared by combining random
numbers of 2D Gaussian with randomly determined amplitude, $\sigma_{x(y)}$,
$\mu_{x(y)}$ and tilt angle. Again, all images were then scaled to $[0,1]$ using
the min-max pixel values of this dataset.

Finally, a Wasserstein Gradient Penalty GAN~\cite{WGAN} was constructed and
trained on images from the real image dataset. As we were interested in the
possibility of producing synthetic data from small training sets, the W-GAN was
trained on only 200 images. This training took about 10 hours on a 12~GB NVIDIA
Tesla K80 GPU, and a synthetic dataset of 1'588 images again scaled in the range $[0,1]$
was produced.

The 8 datasets (6 real data, 2 synthetic) were used to train different
$\beta-$VAEs, from which the encoder circuits were used in the synthetic AWAKE
model (Fig.~\ref{fig:synthetic}) to compare the results.

Both sets of synthetic images are significantly different from the real data,
but could be well reconstructed by the $\beta-$VAE trained on each dataset.
Examples are shown in the Annex, Fig~\ref{fig:vae_example} and \ref{fig:sfig3}. RL tests
were made with each encoder on the synthetic AWAKE model, using the TD3 agent.
It should be noted that the synthetic AWAKE model was based on data taken in
2018, i.e. using a different set of images to those used for the real image
datasets described above.

\begin{figure}[ht]
    \centering
    \includegraphics[width=0.95\linewidth]{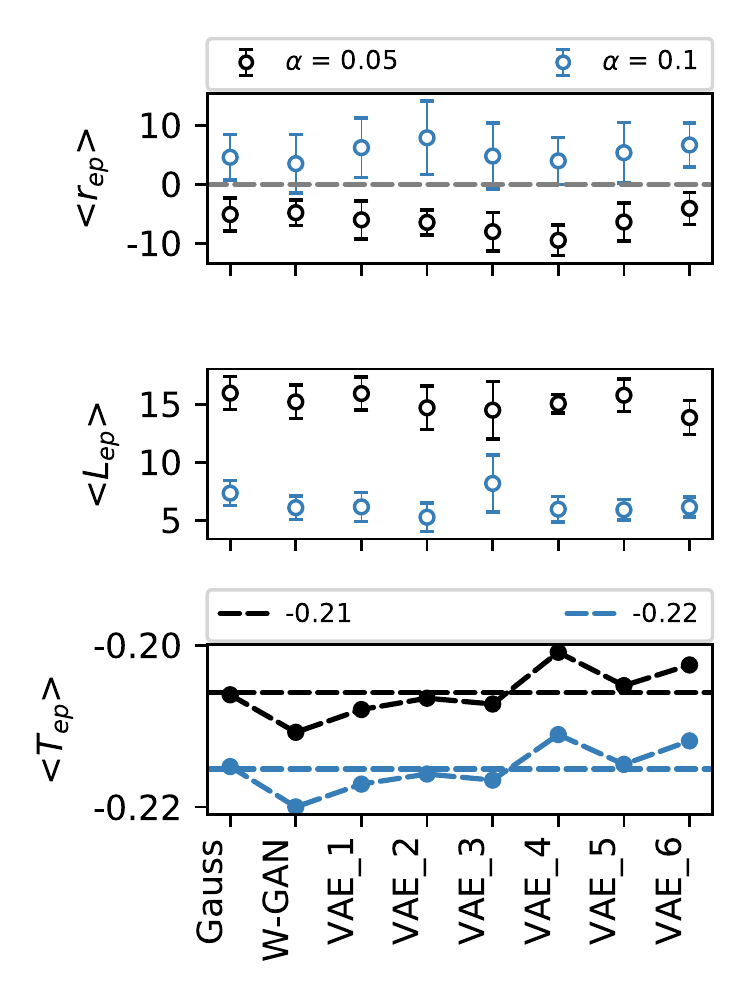}
    \caption{Comparison of six different $ \beta$-VAE, W-GAN and $ \beta$-VAE
    trained on fully fake data, using the synthetic beamline environment
    described in the previous sections. The top plots shows the average total reward
per episode collected by a trained agent when deployed in realistic conditions.
The middle plot shows the average episode length and the bottom one, the target
reward used for the final training after reward dangling. In all these plots,
the average is done over the validation test, assigning an additional positive
reward of 10 if the agent managed to overcome the threshold.
The error bars represent the standard deviation of the results taken over 10
different trained agents.}
    \label{fig:comparison_enc}
\end{figure}

A full comparison of the above described encoders, together with the classic
explicit representation of the state space, is shown in
Fig.~\ref{fig:comparison_enc}, where basically no difference can be seen on the
choice of the encoder type. In the figure, the effect of $ \alpha$ can be
clearly seen, where all the trained agents with slightly larger $ \alpha$
succeed to pass the final reward target, at the cost of a very slightly lower
instantaneous reward.

\section{Results on AWAKE beamline}

\subsection{Numerical optimisation tests }

\begin{figure}[ht]
    \centering
    \includegraphics[width=0.45\textwidth]{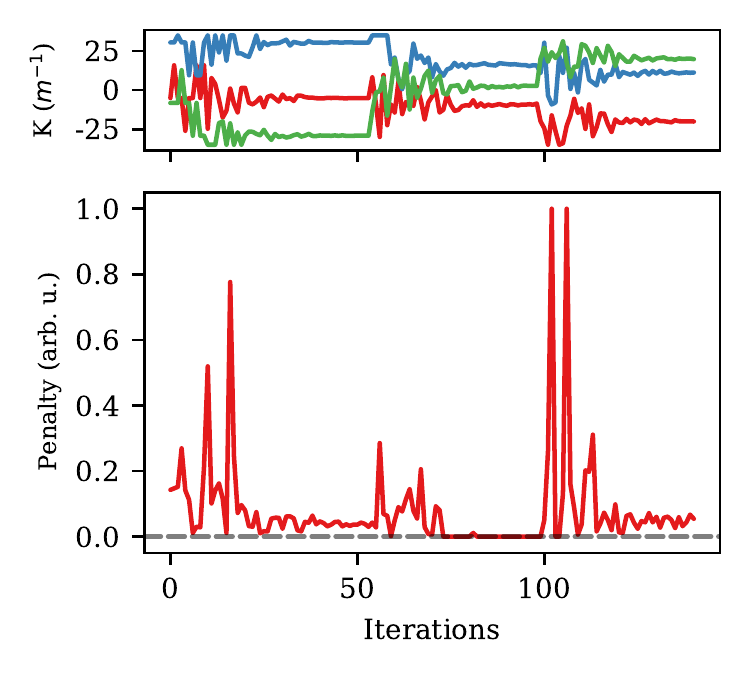}
    \caption{Optimisation result for BOBYQA algorithm using only the first 3 quadrupoles of the
    beamline. The global minimum is represented by the grey dashed line.}
    \label{fig:optmiser_3dof}
\end{figure}

The usage of numerical optimisers has been explored and some of the main
results already published in~\cite{alex_multi}. A large set of algorithms were
tested and most of them showed successful results, although still needing
a large number of iterations (larger than 100 in most cases) and ensuring
that a global minimum search was
in place. An example of a successful optimisation is shown in
Fig.~\ref{fig:optmiser_3dof}. In this example, the target beam size parameters
were obtained using only 3 degrees of freedom, i.e.
the initial quadrupole triplet, but very similar
results were obtained also for 5, as detailed in~\cite{alex_multi}.
All of the experiments succeeded when the
beam size requested was indeed a global minimum for the line. If the requested size
was larger, convergence was not achieved, as the only quadrupoles used in the optimisation
procedure are the initial three and not those responsible for the final focus.
The impact of the initial conditions is nevertheless very significant and
cannot be neglected. As these are not stable, the optimisation procedure needs
to be regularly run to obtain the design beam size at the BTV or plasma cell.

\subsection{Reinforcement learning tests}

For the RL tests the metrics used to measure the performance were the cumulative
reward per episode $r_{ep} = \sum_{i}^{ep}r$, the length of episode $L_{ep}$ in
number of interactions, and the final weighted reward of the last iteration
$r_{ep}$.

\begin{figure}[ht]
    \centering\includegraphics[width=0.4\textwidth]{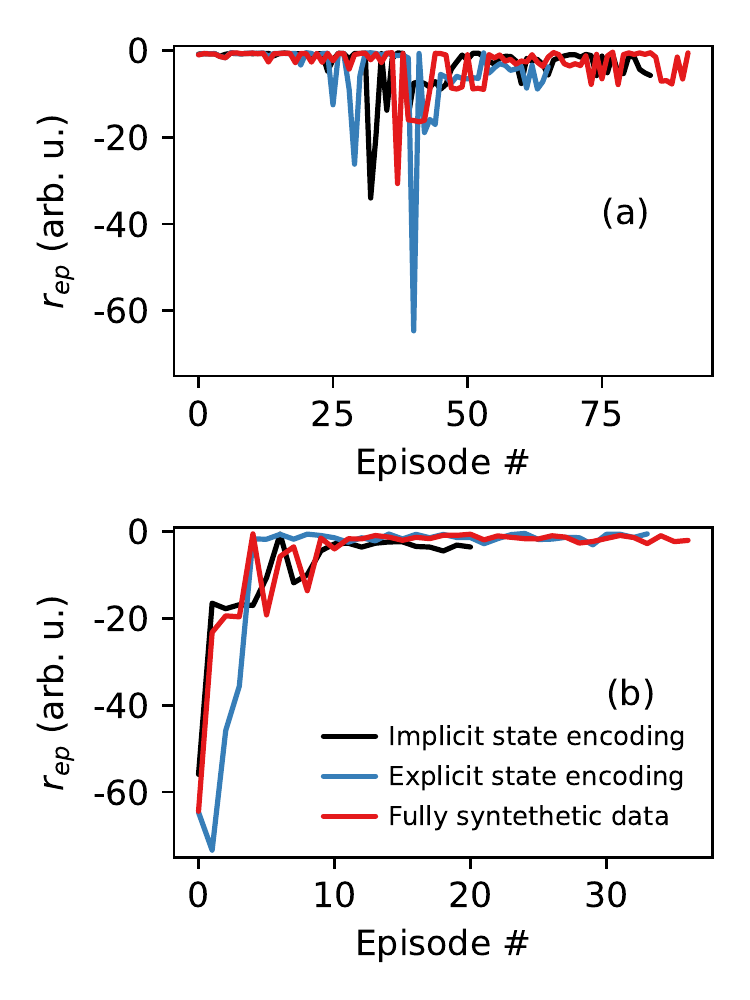}
    \caption{Reward per episode for two-step training process for TD3 RL agent on real
        AWAKE machine. (a) Initial stage of \textit{reward dangling} and (b) actual
        training of TD3 agent with the maximum reward target found.
        The training were performed using three different state representation: implicit state
    encoding with VAE trained on real images, explicit state encoding and implicit state encoding using
    fully synthetic data for VAE training.} \label{fig:real_training}
\end{figure}

\begin{figure}[ht]
    \centering\includegraphics[width=0.4\textwidth]{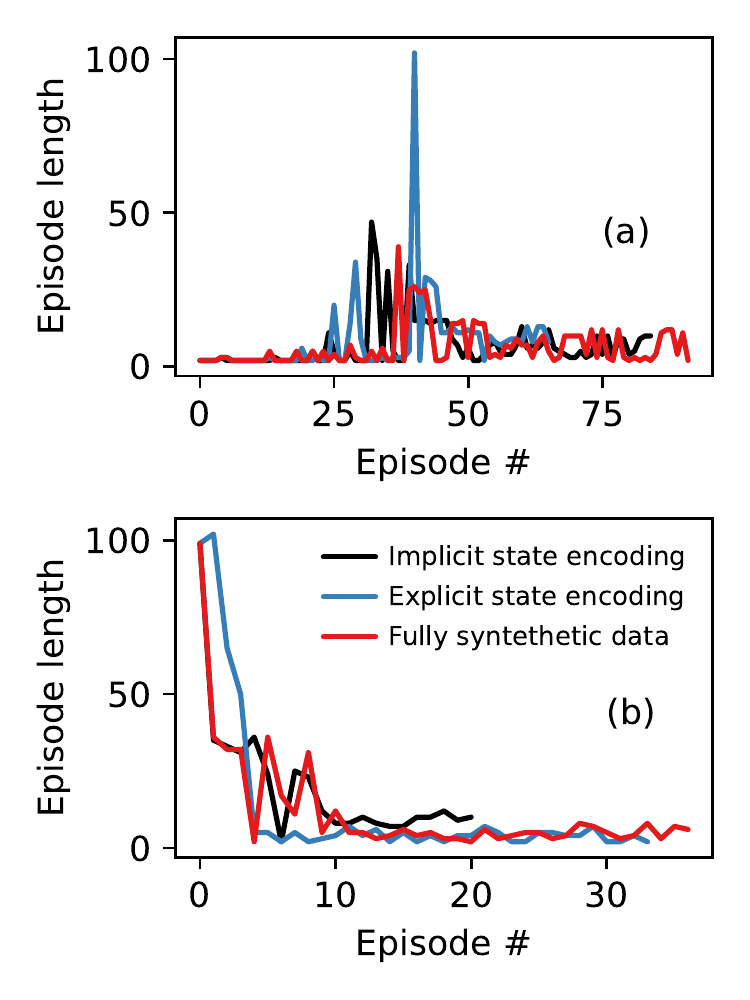}
    \caption{Episode length for two-step training process for TD3 RL agent on real
        AWAKE machine. (a) Initial stage of \textit{reward dangling} and (b) actual
    training of TD3 agent with the maximum reward target found, as for Fig.~\ref{fig:real_training}.}
        \label{fig:training_len}
\end{figure}

After extensive tuning of hyperparameters for different agents and the
development of the reward dangling approach using the synthetic AWAKE model,
tests were made with the real AWAKE machine. The Stable Baselines DDPG RL agent
`TD3' worked reliably and learned the problem dynamics, taking about 350
iterations in each of the two training phases.
The system was then tried with the implicit state extraction, using different
versions of encoder trained on the various real and synthetic datasets described
above. The RL agents also converged in a similar time to the explicit state
versions, showing that the RL state can be successfully auto-encoded in an
unsupervised manner, as shown in Fig.~\ref{fig:real_training} and
Fig.~\ref{fig:training_len}.

Equally importantly, for applications where sample efficiency of the overall
method is important, we demonstrated that encoders trained with fully or
partially (W-GAN) synthetic data were also effective for state encoding. This
is illustrated in Fig.~\ref{fig:real_training} and compared with fully
explicit state description and with an implicit one but trained on real data.
The training time is rather similar in the terms of total machine interactions,
which is about 600 iterations for the first stage where the target is adapted to the
agent performance and about 450 for the actual agent training. The episode length
reached after training for all three different ways of encoding the states is rather
similar and less than 10 machine iterations in all cases.

\begin{figure}[ht]
    \centering
    \includegraphics[width=0.4\textwidth]{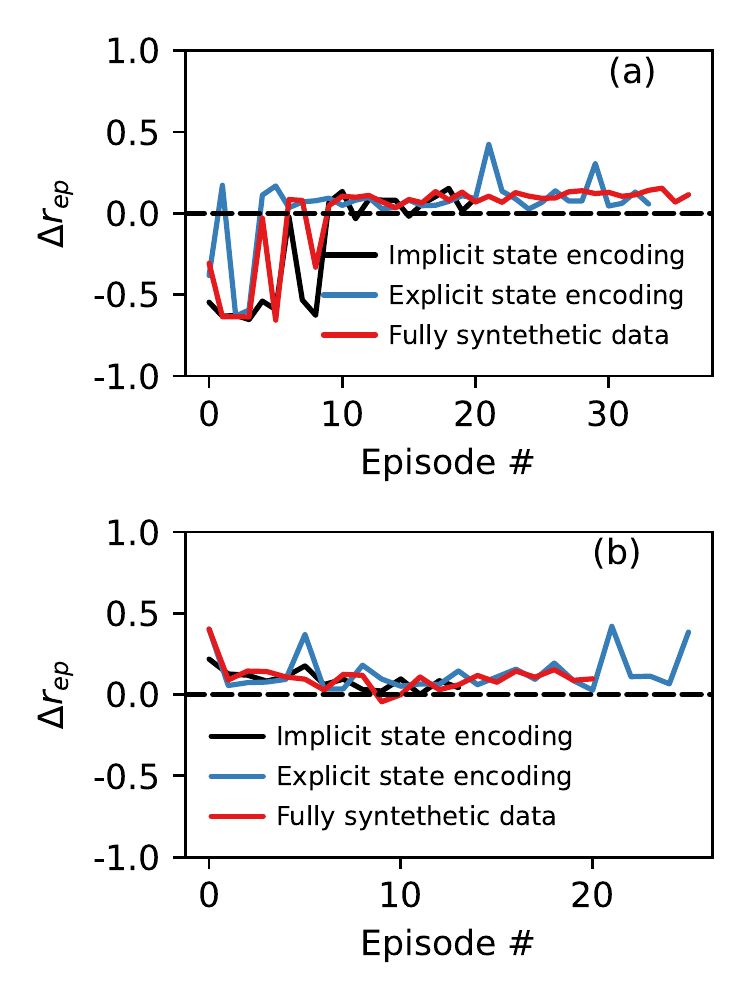}
    \caption{Evolution of $\Delta r_{ep}$ during training (a) and validation (b) for
    three different state encoding. }
    \label{fig:real_validation}
\end{figure}

Another metric to show the evolution of the training (after choosing the target reward
with the \textit{reward dangling} algorithm) of the TD3 agent is $\Delta r_{ep} \equiv
r_e - r_0$, which gives a magnitude of the improvement on the environment made by
the agent after a reset. In Fig.~\ref{fig:real_validation}-a it is clear that after 10
episodes all the agents manage to always improve the performance of the beamline.

The performance of the three different agents trained on the real beamline are
summarised in Fig.~\ref{fig:real_validation}-b. For each agent, the delta reward between
start and end of the episode is plotted as a function of the episode number used to
perform the validation. In this situation the agents trained are free to operate on the
beamline in the context of an episode after the random reset of the actions.

\begin{figure}[ht]
    \centering
    \includegraphics[width=0.5\textwidth]{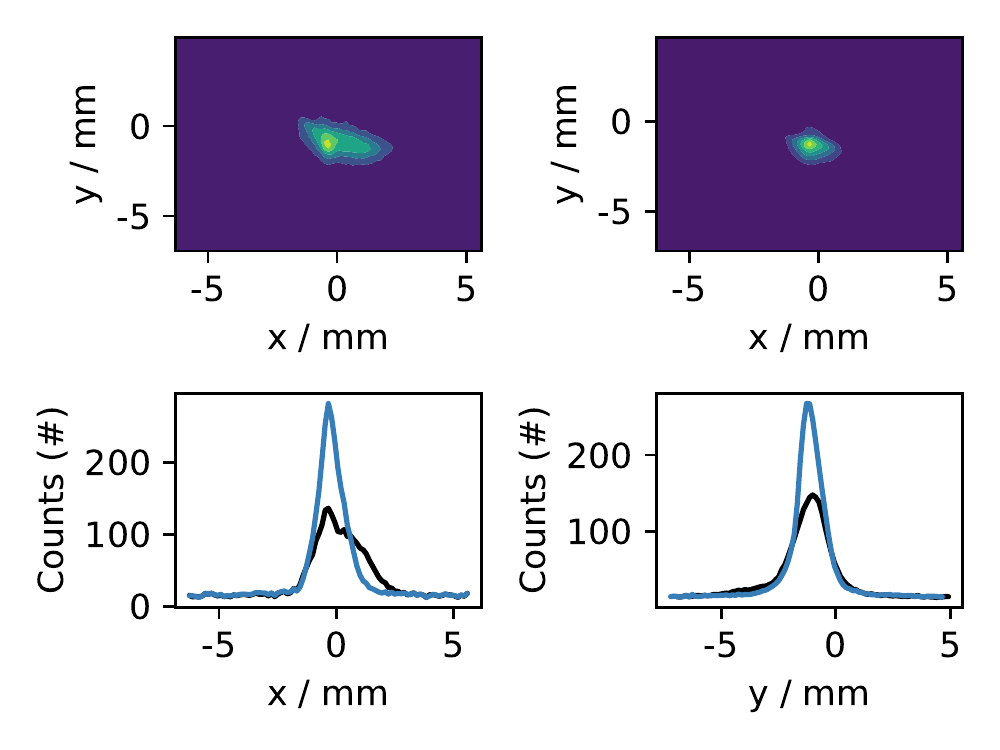}
    \caption{Beam size measurements for one episode after training of TD3 agent using
        implicit state description and VAE trained on synthetic data.
        (Top-left) Measurements at the start of the episode after randomly
        reset the action settings. (Top-right) BTV
    image after the agent has acted is shown. (Below) Horizontal
(black) and vertical (blue) projections of the images above.}
    \label{fig:btv_data}
\end{figure}

The final beam spot obtained was of
very good quality, as shown in the projections plotted in
Fig.~\ref{fig:btv_data}, when using implicit state encoding from a VAE trained on
synthetic data.

The main problem encountered with the RL approach was the longer-term stability
of the beamline. Although the trained agent performed well if tested during a
few hours of the training, the results were less good when tested some weeks or
months later. In these cases, the trained agent failed to converge, indicating
that the problem dynamics had shifted outside of the valid training data space.
The contributing
factors are likely to be the `hidden' action variables which change the beam
spot distribution and hence the encoded state - this is not surprising given the
adjustments made to the source including the laser power, alignment,
synchronisation and RF phasing which are all empirically adjusted before a new
run, or indeed at the start of each day during a running period.

\section{Discussion}

\subsection{Usage of RL agents in operation}%
\label{sub:usage_of_rl_agents_in_operation}

The training of RL agents using implicit or explicit state encoding proved to be
successful regardless on the encoding chosen and on the initial state of the
system - it was in fact possible to train a large number of RL agents on different
days, where the source states (unknown to the agents) were changing either on
purpose or randomly. The main problem though, was that trained agents were very
difficult to reuse days after their training, showing very poor performance.
We believe that this behaviour is to be attributed to the unknown states
of the source and basically to the change of the mapping between actors, states and
reward that follows. A clear solution would be to include more state dimensions
describing the source and the possible parameter configuration, but unfortunately the
beam time available was not sufficient to test also this state parameterisation.

These results are not suitable for operational deployment of the trained RL agents,
but clearly showed their potential. For this reason, it was decided to rely on
numerical optimisers to provide the required beam quality to the experiment and
continue the studies on the usage of RL agents with a larger state description.

\subsection{Performance reach}%
\label{sub:performance_reach}

Numerical optimisers could take up to a few hundred of iterations to achieve a
suitable beam configuration for the experiment, strongly depending on the initial
conditions of the optimisation. RL agents, instead, could perform this task in just
a handful of action steps, if the mapping at training time is preserved. Clearly this
would mean a huge improvement in the usage of machine time, although work it is
still needed to include the hidden states that are causing the variation of the mapping
in time.

The full process is done today all manually and using linear optics approximations to
treat the line initial condition changes. Already the deployed environment and the
tested numerical optimisers would provide a speed up in setting up time and possibly
more reproducible conditions for the experiment. Studies are still ongoing to fully
deploy operationally this method.

A possible extension to the manual optimiser launching is to use model-free
adaptive feedback systems, like extremum seeking~\cite{extremum_seeking}, to
maintain the optimal value found via numerical optimisation even after drift of
the settings. This is of course only possible in case the drift are slow with
respect to the probing frequency.

\subsection{Applicability to other systems and outlook}%
\label{sub:applicability_to_other_systems}

The tools and methodologies presented in this paper are rather generic and the
application to the AWAKE transfer line case could be seen as a first proof-of-principle.
The unsupervised state encoding via auto-encoders could be used in many other systems and domains
of accelerators. Also, the methodology presented to assess the optimal reward threshold for
RL agents training can be equally applied to any other RL training case were the
episode termination threshold strongly drives the training speed and success rate. For example,
this is under study for the training of RL agents to automatically extract
information from Schottky spectra in the CERN Low Energy Ion Ring
(LEIR)~\cite{shottky-vision}.

The full methodology as described could also have other applications in transfer lines with
very sensitive final focus. Depending on the available instrumentation, the
same basic methodology, but, for instance, using multiple screens for more
accurate optics estimation, could be envisaged. The technique presented in this
paper could be applied almost entirely, basically changing only the source of
image.

Looking at further development, the access to multiple beam observations could
open the way to a full phase-space reconstruction using lower projection needed
than classic tomography. This would allow a more detailed state description and
hence a much more robust system to unknown variables and drifts.

\section{Conclusions}

The feasibility of automatic optimisation of the AWAKE e$^-$ beamline brightness
based on machine learning techniques has been successfully explored using a
variety of approaches. The performance of suitable numerical optimisers was
demonstrated, despite the non-convex nature of the overall problem. A number of
other useful techniques have been developed and tested, including successful RL
agent training, a generative synthetic model using $\beta$-VAE for offline
testing and hyperparameter optimisation, the use of implicit unsupervised RL
state encoding with a $\beta$-VAE encoder based on computer vision, the training
of these encoder networks with fully synthetic data and the development of
automatic reward target value setting for RL episode termination.

The limitations of RL were also reached for this specific configuration, where
the variations from the e$^-$ source meant that the trained RL agents
could not reproducibly be deployed over long time scales. The work showed the
importance of capturing all the generative factors in the observation of the RL
state space, as well as including all the corresponding actions which are used
to correct the performance. Given the high repetition rate of the AWAKE e$^-$
beamline, the simpler optimisation approach is the one which will be deployed
operationally. Nevertheless, the RL paradigm with a learned response remains
relevant for applications like injection into LHC, where each sample is much
more expensive to take, and the repetition rate is orders of magnitude lower.

\section{Acknowledgements}

The stimulating real-world environment provided by the CERN ``Machine Learning
Coffee'' forum \cite{MLcoffee} proved invaluable for presenting and discussing
progress and results. We also benefited greatly from the dedication of the
entire AWAKE project team, the CERN Equipment groups and the SPS operations
team. Particular thanks go to E.~Gschwendtner for her continued enthusiastic
support of these activities, as well as to L.~Verra and S.~Dobert for keeping the
beamline running when we needed it.


\bibliographystyle{ieeetr}

\clearpage \pagebreak

\onecolumngrid

\appendix

\section{Synthetic images for different $\mathbb{Z}$ dimensionalities} Random
samples of real, VAE recovered and fully synthetic 128x128 pixel images for
different Z dimensionalities. The images in the top rows are the original BTV
measurements, those in the lower rows have been generated through the full
synthetic AWAKE model from the labelled action values, using the predictor and
decoder networks.

\begin{figure*}[h]
\centering\includegraphics[width=0.9\textwidth]{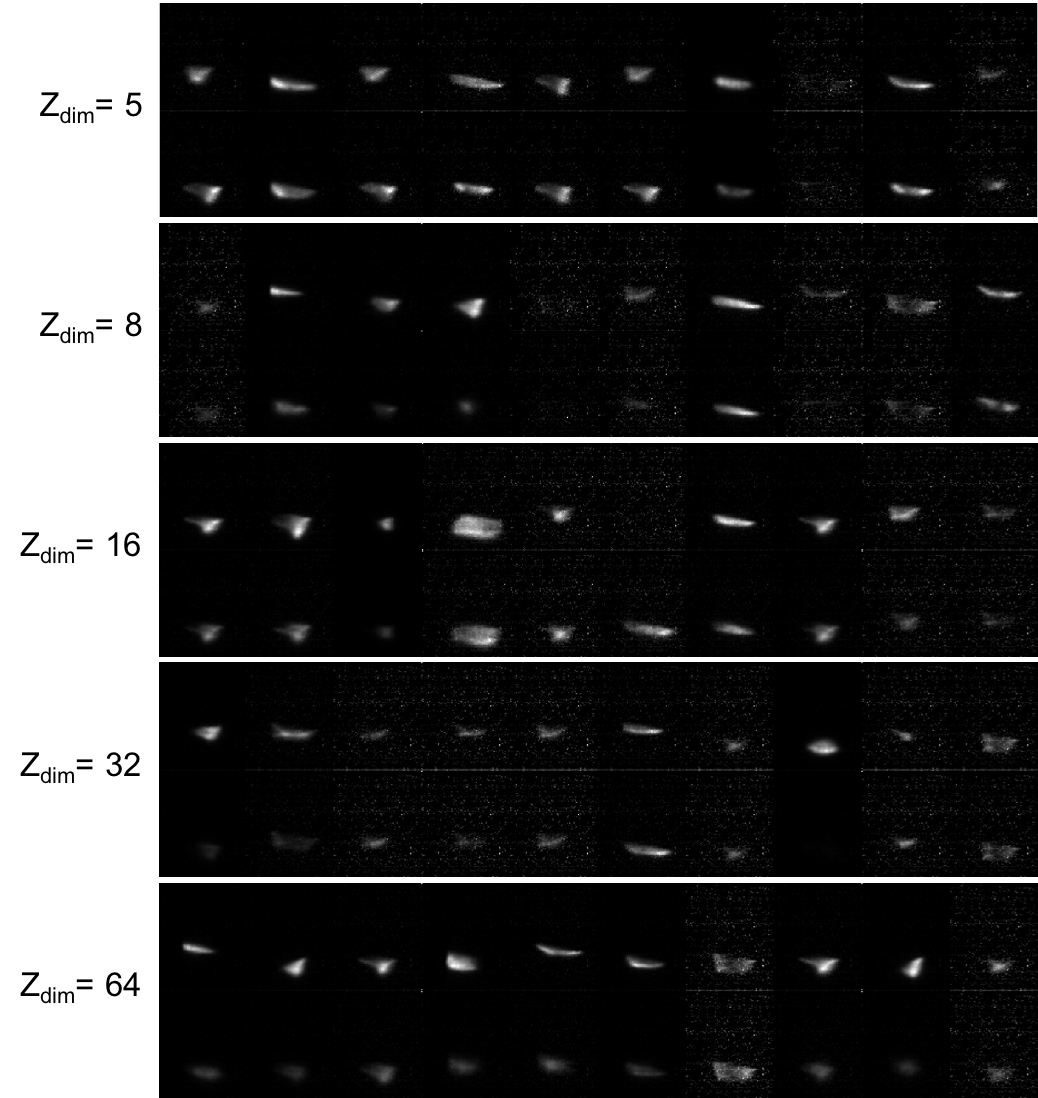}
\caption{Example of real (top rows) and images produced by the VAE (bottom
rows) for different dimension of the latent space.}
\label{fig:vae_images_z}
\end{figure*}

\clearpage \pagebreak

\section{Real and synthetic images for VAE state encoder training}

Random samples of 128x128 pixel images used to train VAE for image encoding (top
rows) and recovered VAE images (bottom rows) for the real and two synthetic
datasets.

\begin{figure*}[h]
    \begin{subfigure}{.9\textwidth} \centering
    \includegraphics[width=.99\linewidth]{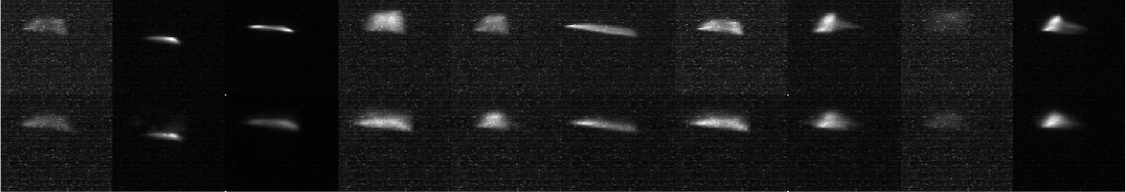} \caption{Real
    AWAKE BTV images (top) and $\beta$-VAE reconstruction (bottom).} \vspace{12px}
    \label{fig:zreconstruction}
    \end{subfigure}

    \begin{subfigure}{.9\textwidth} \centering
    \includegraphics[width=.99\linewidth]{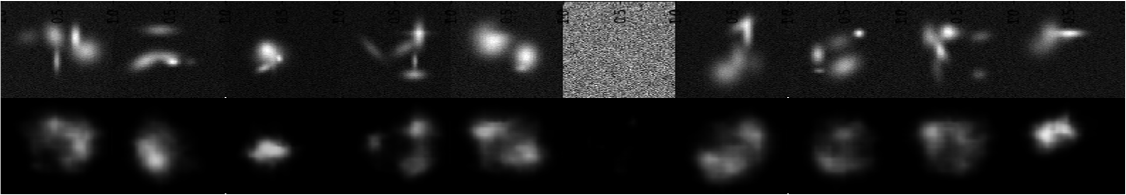}
    \caption{Synthetic images from analytical superimposed Gaussians (top) and
    $\beta$-VAE reconstruction (bottom).} \vspace{12px} \label{fig:vae_example}
    \end{subfigure}

    \begin{subfigure}{.9\textwidth} \centering
        \includegraphics[width=.99\linewidth]{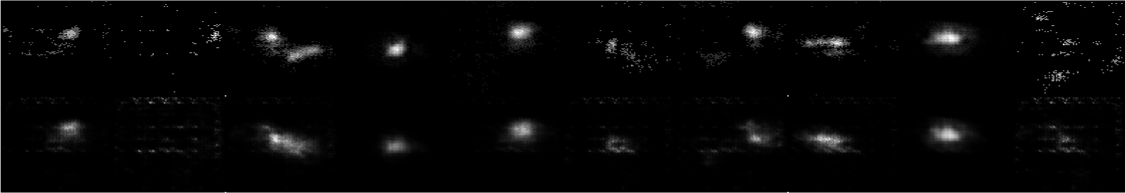}
        \caption{Synthetic images from Wasserstein GP-GAN (top) and $\beta$-VAE
        reconstruction (bottom).} \label{fig:sfig3}
    \end{subfigure}
\caption{Example of real images, analytically and W-GAN produced images are shown on
    top rows. Examples of images reconstructed with the VAE are instead shown
on bottom rows.}
\end{figure*}

\end{document}